# A Framework Based on Graph Cellular Automata for Similarity Evaluation in Urban Spatial Networks

Peiru Wu, Maojun Zhai, Lingzhu Zhang

College of Architecture and Urban Planning, Tongji University, Shanghai 200092, China

wupeiru@tongji.edu.cn (P.W.); zhaimj@tongji.edu.cn (M.Z.); zhanglz@tongji.edu.cn (L.Z.)

## Abstract

Measuring similarity in urban spatial networks is key to understanding cities as complex systems. Yet most existing methods are not tailored for spatial networks and struggle to differentiate them effectively. We propose GCA-Sim, a similarity-evaluation framework based on graph cellular automata. Each submodel measures similarity by the divergence between value distributions recorded at multiple stages of an information evolution process. We find that some propagation rules magnify differences among network signals; we call this "network resonance." With an improved differentiable logic-gate network, we learn several submodels that induce network resonance. We evaluate similarity through clustering performance on fifty city-level and fifty district-level road networks. The submodels in this framework outperform existing methods, with Silhouette scores above 0.9. Using the best submodel, we further observe that planning-led street networks are less internally homogeneous than organically grown ones; morphological categories from different domains contribute with comparable importance; and degree, as a basic topological signal, becomes increasingly aligned with land value and related variables over iterations.

## Keywords

## 1. Introduction

Urban spatial networks, such as road, ecological, and pedestrian networks, link urban space and carry multidimensional information. They offer a clear entry point for studying the city as a complex system (Barthélemy, 2011). Similarity is also a core topic in complex systems (Tsitsulin et al., 2018; Tantardini et al., 2019). In urban science, finding common macro patterns emerging from diverse micro configurations helps reveal deep regularities of urban development and informs evidence-based policy and design (Batty, 2008; Louf & Barthélemy, 2014). In smart-city practice, similarity metrics guide model optimization in machine learning and support feature recognition, classification, and generative design, with broad prospects in digital twin applications (Bellet et al., 2013; Kulis, 2013; Batty, 2018). However, many existing approaches rely on empirical labels or static snapshots and are not tailored to spatial networks, which limits their power to identify similarity.

First, experience-driven analyses have inherent limits and cannot uncover the prior knowledge embedded in similarity (Louf & Barthélemy, 2014). Classic urban studies judged similarity by visual form and grouped street patterns by expert reading, for example Muratori's (1959) work on Venice and Conzen's (1960) typology of street forms. Modern learning can estimate the probability that a network belongs to a label, but the mapping between samples and labels is still defined by prior experience, so the method's validity is bounded by the labels' validity (Wang et al., 2024). We therefore use label-free clustering performance to reflect a model's capacity to identify similarity.

Second, city-scale road networks often have $10^4$–$10^6$ nodes, and many methods fail to balance complexity and accuracy. For example, Laplacian spectral distances need full eigendecomposition with worst-case time complexity $O(n^3)$; doubling nodes leads to about eight times the computation (Golub & Van Loan, 2013), so typical usable sizes are about $10^3$ nodes or fewer. We use node neighborhoods as the computational unit, avoid global computation, and keep complexity near linear to fit large spatial networks.

Third, physical constraints make spatial networks more alike in basic distributions. In contrast to social or citation networks, the degree at an intersection is usually a small integer, so methods that rely on degree distributions, such as degreeJSD, struggle to distinguish networks that look similar on the surface (Carpi et al., 2011). Yet small local changes can cause large macroscopic effects, for example a single road closure that triggers congestion. We therefore use dynamic information to pass signals and amplify effects, which makes similarity easier to detect.

Fourth, spatial network information shows spatial dependence and spatial heterogeneity (Tobler, 1970, 2004). Many similarity measures compare a single global distribution, which flattens spatial attributes and loses geographic structure (Schieber et al., 2017; Guo et al., 2013). We confine each node's influence to a multi-hop neighborhood within a fixed number of iterations and record multi-stage slices of information to preserve and transform spatial attributes.

Taken together, the core bottleneck in evaluating similarity is how fully we use urban information. Although similarity depends on context (Goodman, 1972), it ultimately comes from the information carried by the two networks. Traditional methods often capture only a single slice of complex spatiotemporal interactions. A better approach should rely on dynamic information. The dynamics do not need to mimic real interactions; as attributes propagate, they also reveal structure, so even simple processes can fuse information (Holme & Saramäki, 2012; Coifman & Lafon, 2006; Masuda et al., 2017). Some propagation rules help assimilate or differentiate information and thus improve clustering. We call this "network resonance." Our goal is to find propagation rules that drive resonance and, by using information more fully, reveal latent similarity in spatial networks.

This study makes three contributions. First, it proposes a GCA-based framework for similarity analysis in urban spatial networks. Each submodel evaluates similarity on multi-stage slices of an information evolution process, works without labels, and fits key properties of spatial networks. Compared with existing models, methods under this framework show large gains in similarity recognition and clustering and perform well in practical settings. Second, it introduces an improved differentiable logic-gate network that removes unnecessary numerical constraints and simplifies operations, which helps identify interpretable rules behind complex dynamics. Third, it reveals the phenomenon of network resonance and explores its mechanisms.

# 2. Literature review

## 2.1. Methods for Evaluating Graph Similarity

Similarity in urban spatial networks reduces to similarity between graphs. Early methods compare node or edge sets directly, such as Vertex/Edge overlap defined by the share of common edges between two graphs (Papadimitriou et al., 2010). These methods assume a one-to-one correspondence between node sets (Shvydun, 2023), which does not hold for real spatial networks.

To address this limit, some approaches compare global feature distributions and can handle unlabeled graphs with different numbers of nodes (Shvydun, 2023). Examples include spectral distances such as the Ipsen–Mikhailov distance, motif-based methods such as Graphlet Degree Distribution, and the state of the art Gromov–Wasserstein (GW) framework (Ipsen & Mikhailov, 2002; Pržulj, 2007; Mémoli, 2011; Peyré et al., 2016). As noted earlier, these methods often have high computational cost.

Methods suited to urban networks must scale across sizes while balancing efficiency and accuracy. Representative options include divergence-based comparisons such as NetSimile (Berlingerio et al., 2012) and kernel methods such as the Weisfeiler–Lehman (WL) graph kernel (Shervashidze et al., 2011). Yet most of these approaches apply preset statistics to static snapshots, so they struggle to capture the nonlinear interactions that arise as information evolves in space and time.

Recent machine-learning methods on graphs, including graph cellular automata (GCA) and graph neural networks (GNN), have been used for similarity tasks. They preserve network topology and spatial attributes, and they can fuse multi-hop neighborhoods with node features to capture richer structure (Scarselli et al., 2009; Kipf & Welling, 2017; Hamilton et al., 2017). Existing studies that apply such methods to spatial networks often focus on isomorphism or non-isomorphism within a single network (Xue et al., 2022; Tian et al., 2025) rather than measuring similarity between two spatial networks directly.

## 2.2. Graph Cellular Automata

Cellular automata (CA) are a key tool for studying the evolution of complex systems. A CA specifies cell states, a neighborhood, and an update rule. Each cell updates synchronously based on its local neighborhood under a common rule. Conway's Game of Life is a classic example in which simple local interactions on a two-dimensional grid yield complex global patterns (Gardner, 1970). Traditional CA operate on regular lattices, while graph cellular automata (GCA) extend the architecture to arbitrary graphs. GCA retain rule-driven updates and neighborhood propagation and support state evolution on any graph (O'Sullivan, 2001; Grattarola et al., 2021). Since their introduction, GCA have been applied to many tasks on urban spatial networks, including traffic flow simulation (Małecki, 2017), epidemic modeling (Martínez et al., 2013), and community detection (Bagnoli et al., 2012). We therefore adopt GCA as the base architecture for information exchange in urban spatial networks.

## 2.3. Differentiable Logic-Gate Networks

To discover rules that drive complex dynamics, some studies replace hand-coded CA rules with neural networks that learn update rules automatically, an idea known as neural cellular automata (NCA) (Mordvintsev et al., 2020). Conventional neural networks, however, are hard to interpret. To improve interpretability, recent work replaces standard neurons with differentiable logic gates (Petersen et al., 2022, 2024; Benamira et al., 2024). Logic gates are the basic units of digital circuits; their outputs are binary and cannot be optimized by gradient descent. Differentiable logic gates map Boolean logic into continuous, differentiable functions so they can substitute for neurons in learning. A representative example is DiffLogic-CA by Miotti et al. (2025), which combines differentiable logic gates with NCA to recover interpretable rules for complex phenomena, including the local update rules of the Game of Life. This line of work has limits for spatial networks: it mainly targets regular grids; it restricts inputs and outputs to the range [0, 1], which fits many urban signals poorly; and its Boolean foundation makes basic arithmetic operations hard to express. We therefore modify differentiable logic gates to keep their high interpretability while overcoming these limits.

# 3 Data and methodology

## 3.1. Dataset

We obtained urban road networks (URNs) for 146 cities from OpenStreetMap (OSM) (Boeing, 2017). The split is 64 for training, 32 for validation, and 50 for analysis. City selection balanced geographic coverage and OSM data availability. Fig. 1 shows the spatial distribution of the samples.

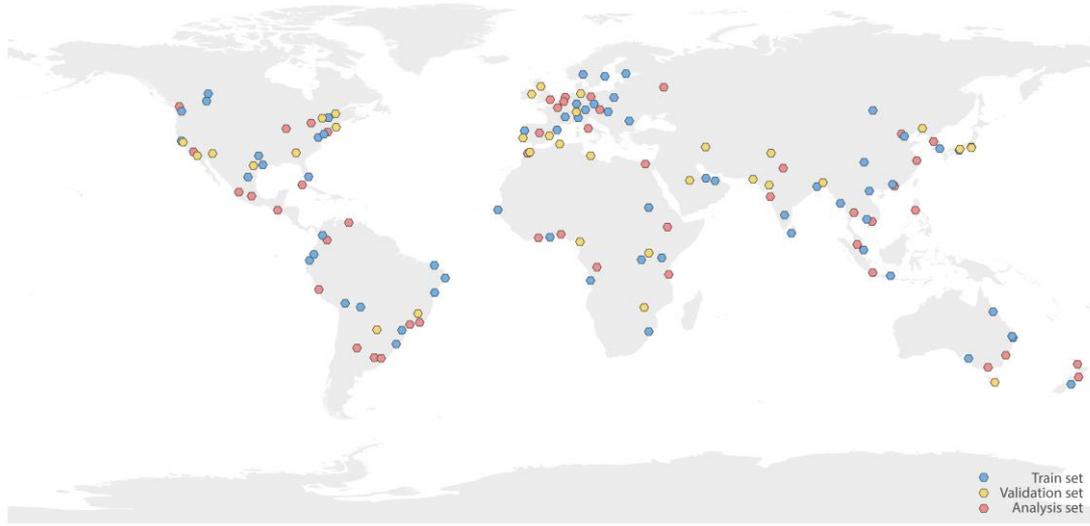

Fig. 1. Distribution of urban spatial network data.

To reduce computation, we trained and validated on district-level road networks extracted from the above sets. The training part has four groups and the validation part has two groups. Each group contains 32 networks.

## 3.2. Differentiable Logic-Gate Design

We modify differentiable logic gates by replacing the operator set with basic arithmetic and routing operations. Max and Min provide the essential nonlinearity (Goodfellow et al., 2013). These simple operators avoid issues such as gradient explosion that can arise with complex functions, improve interpretability, and extend the domain to the entire real line, which avoids information loss due to normalization. During training, each gate learns a probability mix of these base operators through a softmax and then outputs the most probable operator to form a deterministic differentiable-gate network. The gates used in this study are listed in Table 1.

| Operation | Expression | Symbol |
|---|---|---|
| Add | a + b | |
| Subtract | a - b | |
| Reverse Subtract | b - a | |
| Max | max(a, b) | |
| Min | min(a, b) | |
| Pass A | a | |
| Pass B | b | |
| Negate A | -a | |
| Negate B | -b | |

Table 1. Differentiable logic gates used in this study.

## 3.3. GCA-Sim: A Similarity Evaluation Framework Based on Graph Cellular Automata

We propose a GCA-based framework for similarity evaluation, which sits above any specific model. Submodels under this framework share three traits: they use the node neighborhood as the basic unit of computation, iterate network information according to a rule, and compare multi-order value distributions of the evolving information to measure similarity.

Fig. 2 shows the architecture. Three modules suffice to evaluate similarity. First, we compute and store the information needed for evaluation. Second, we iterate that information under a chosen rule. Third, we record the value distribution of node states at each iteration and compare the distributions between two networks to measure similarity. To discover new submodels within the framework, we add a fourth module that loops over the first three, clusters networks using the similarity index as a distance, and searches for submodels with strong clustering metrics.

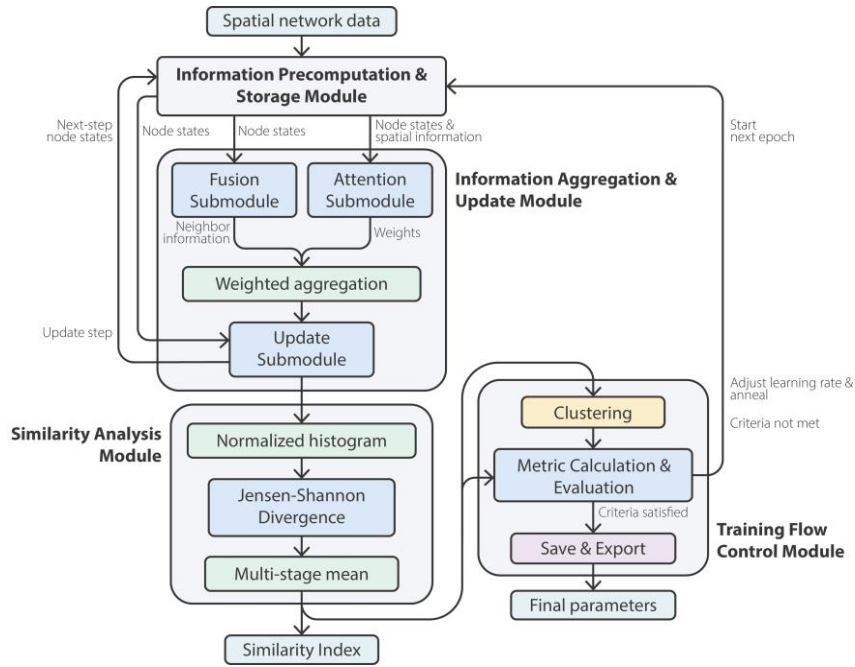

Fig. 2. Architecture of the GCA-based similarity evaluation framework.

## 3.3.1. Information Precomputation and Storage Module

This module computes and stores the initial data for similarity. Each node has two storage slots. The first stores the current state $s^0$. We set the current state to the node's degree because degree is a basic descriptor of network topology and is widely used in network similarity, so submodels capture topological similarity as it appears in the evolution of the degree distribution (Newman, 2008; Berlingerio et al., 2012). The second stores spatial information. For each outgoing edge we record two values: normalized distance $d$ and normalized angle $\theta$. We compute distance in meters in geographic coordinates using the fast geodesic Haversine formula (Cotter, 1974), then normalize by the maximum over the node's incident edges. For angle, we take the sum of the angles between the edge and its left and right neighbor edges, then normalize by 360°.

## 3.3.2. Information Aggregation and Update Module

When used only for evaluation, this module can run with a fixed propagation rule. To search for new submodels, we replace the fixed rule with a learnable network of differentiable logic gates. The basic unit is a node pair. The Fusion and Attention submodules run in parallel to compute the message and the weight for each pair, then aggregate them at the center node. The Update submodule combines the pairwise information to update the center node's state. This design accommodates variable neighborhood sizes in urban spatial networks (O'Sullivan, 2001; Grattarola et al., 2021). Fig. 3 shows the structure.

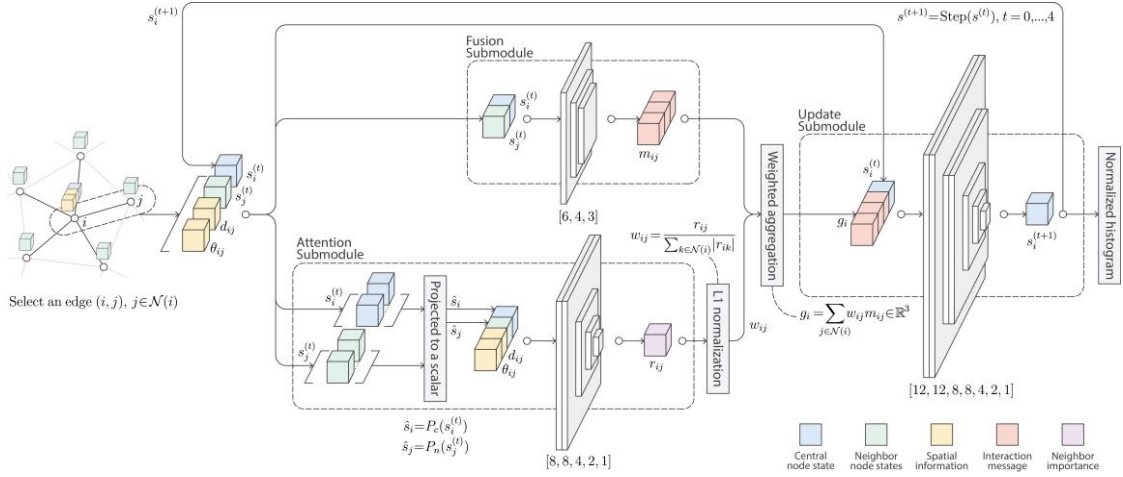

Fig. 3. Structure of the information aggregation and update module.

The Fusion submodule captures interaction between two nodes. It takes only the two current states and does not use spatial features; it serves to integrate information. The Attention submodule measures how important a neighbor is in the current round. For each center node, we apply L1 normalization to the attention outputs over all outgoing edges to obtain edge weights (Koohpayegani & Pirsiavash, 2024). We then take a weighted sum of the Fusion outputs to get the aggregated neighbor message. This design is permutation invariant to the order of neighbors and adapts to nodes with different degrees (Zaheer et al., 2017; Hamilton et al., 2017). Finally, the Update submodule receives the neighbor-set message and the node's current state, computes the next state, overwrites the old state, and advances the iteration. As the module playing the central role in information exchange, the Update submodule is designed to have a correspondingly larger network size.

### 3.3.3. Similarity Analysis Module

We define similarity between urban spatial networks as the mean similarity across the normalized distributions recorded at successive stages of an information evolution process. Prior work shows that if two objects share the same heat-diffusion response at all time scales, they are structurally equivalent (Sun et al., 2009). We draw on this idea and use multiple temporal slices to reflect differences induced by information propagation, which is fundamentally different from comparing a single distribution.

The computation proceeds as follows. For each network, we run five iterations of the state-update process. At every iteration we record the distribution of node states, normalize it, and map it to a differentiable histogram so it can support gradient backpropagation. We then measure the gap between two distributions with the Jensen–Shannon divergence (JSD), which is bounded and symmetric (Lin, 1991; Endres & Schindelin, 2003):

$$JSD(P|Q) = \frac{1}{2} D_{KL}\left(P \middle| \frac{P+Q}{2}\right) + \frac{1}{2} D_{KL}\left(Q \middle| \frac{P+Q}{2}\right) \quad (1)$$

Here $P$ and $Q$ are two distributions and $D_{KL}$ is the Kullback–Leibler divergence. We take the mean JSD across iterations as the distance between two networks:

$$D_{\text{avg}}(P, Q) = \frac{1}{T} \sum_{t=1}^{T} JSD(P_t|Q_t) \quad (2)$$

where a larger $D_{\text{avg}}$ indicates greater dissimilarity and $T = 5$ in this study.

### 3.3.4. Training Flow Control Module

The training pipeline has two phases: random exploration and fine-tuning.

To avoid poor local optima, we begin with a Monte Carlo–style randomized initialization (Metropolis & Ulam, 1949). Each gate unit outputs probabilities over nine candidate gates in parallel each round, and a randomly chosen candidate acts as the dominant operator. When the average validation metrics across groups reach preset thresholds, the model enters the fine-tuning phase with a higher target.

Fine-tuning adjusts the differentiable-gate network and tests robustness, since well-generalizing solutions should not degrade under small perturbations (Hochreiter & Schmidhuber, 1997; Keskar et al., 2017; Li et al., 2018). We iterate over the training groups, training each group for ten epochs. A weighted multi-term loss guides optimization:

$$\mathcal{L} = \alpha \mathcal{L}_{\text{sil}} + \beta \mathcal{L}_{\text{reward}} + \gamma \mathcal{L}_{\text{margin}} + \delta \mathcal{L}_{\text{ent}} \quad (3)$$

The core term $\mathcal{L}_{\text{sil}}$ measures the distance of the Soft-Silhouette metric to 1. Values closer to 1 indicate tighter clusters and clearer separation, meaning the current propagation rule better reveals similarity (Rousseeuw, 1987; Campello & Hruschka, 2006; Vardakas et al., 2024). The structural-hardening term $\mathcal{L}_{\text{hard}}$ encourages each node to commit to a single gate. The margin term $\mathcal{L}_{\text{margin}}$ prevents concurrent shrinkage of inter- and intra-cluster distances. The cluster-size entropy term $\mathcal{L}_{\text{entropy}}$ avoids extreme balance or collapse to one cluster (Krause et al., 2010).

At the end of each epoch we form a deterministic gate network by choosing, for every node, the gate with the highest probability. We then compute similarities on the validation set, run hierarchical clustering (Johnson, 1967), select the partition that optimizes the clustering metrics, and record the results. Training stops and the model is saved once the metrics are satisfactory.

## 3.4. Validation of Clustering Ability

We compare our method with representative approaches well-suited to spatial networks:

Distribution-based: degreeJSD converts node degrees to probability distributions and measures their difference by JSD (Carpi et al., 2011).

Node-signature summaries: NetSimile extracts local statistics per node and aggregates them into a fixed-length graph signature; distances between signatures give graph similarity (Berlingerio et al., 2012). NetLSD (improved) uses the trace of the heat kernel of the graph Laplacian across time scales as a spectral signature and measures distances between the resulting sequences (Tsitsulin et al., 2018).

Structure-based: DeltaCon_0 approximates node affinities via fast belief propagation and compares affinity matrices (Koutra et al., 2013). Network Portrait Divergence (NPD) encodes multi-ring node counts into a "network portrait" matrix and compares portraits with information-theoretic divergence (Bagrow & Bollt, 2019).

Graph kernel: the Weisfeiler–Lehman subtree kernel captures and counts subtree patterns through iterative label refinement and compares graphs by the kernel inner product (Shervashidze et al., 2011).

We also include a simplified method under our framework, LGCA-Sim, which uses the random-walk graph Laplacian as the iteration rule. The node update is the current state minus the mean of neighbor states (von Luxburg, 2007).

We evaluate models with label-free internal clustering metrics. Besides Silhouette and Soft-Silhouette, we report the Calinski–Harabasz index (CH), the ratio of between- to within-cluster variance where larger is better (Caliński & Harabasz, 1974), and the Davies–Bouldin index (DB), where smaller values indicate clearer separation (Davies & Bouldin, 1979). We use hierarchical clustering and select the number of clusters by a joint rank based on Silhouette and DB.

## 3.5. Selecting the best submodel and application design

To avoid mixing models driven by different propagation rules and thus weakening interpretability, we select the best submodel by a joint rank of the Silhouette score and the number of clusters. In our experiments these two quantities are negatively correlated, so the joint rank reflects a model's capacity to capture both the quality and the quantity of latent patterns.

We test the best submodel in three application settings.

First, we measure similarity for 50 cities and 50 districts worldwide and evaluate internal consistency at the city scale. Following Tian et al. (2025), we take a 20 km by 20 km window centered at the network centroid and split it into 1 km by 1 km subgraphs. Based on pairwise similarity among subgraphs, we define the consistency index for city $c$ as

$$\text{IC}(c) = \frac{2}{K(K-1)} \sum_{1 \leq i < j \leq K} \left(1 - \frac{D_{ij}}{\ln 2}\right) \tag{4}$$

where $K$ is the number of subgraphs and $D_{ij}$ is the similarity between subgraphs $i$ and $j$.

Second, we examine the similarity and continuity of urban texture within cities, using Beijing and Shanghai as examples. We compute similarities among central districts in each city and assess continuity by the similarity between adjacent districts.

Third, to test whether new properties emerge during propagation, we use Shanghai to compute Spearman correlations over ten iterations between node states and three sets of external variables

(Spearman, 1904). The first is the mean length of outgoing edges per node. Together with degree, it reflects one-hop neighborhood information and serves as a baseline against the other variables. The second is a set of space syntax measures that have known links to social phenomena (Hillier & Hanson, 1984; Hillier et al., 1993; Hillier, 1996) and that summarize neighborhood information at different radii. We use the sDNA toolbox in ArcGIS (Cooper & Chiaradia, 2020) and the Hybrid metric that combines metric distance and angular change with a 1:1 weight (Hillier & Iida, 2005; Turner, 2007; Zhang & Chiaradia, 2022) to compute MHD and BtH at multiple radii (in meters) as proxies for closeness and betweenness centrality. We then average values over outgoing edges to the center node. The third is land value, using a 100 m resolution surface for Shanghai (Wu et al., 2025).

## 4. Results

### 4.1. Comparison of Clustering Ability Across Models

For methods with randomness, we run ten trials and report means and standard deviations. Tables 2 and 3 summarize the results.

| Methods | Silhouette | Soft-Silhouette | CH | DB | Clusters |
|---|---|---|---|---|---|
| degreeJSD | 0.323 | 0.645 | 16.711 | 0.557 | 5 |
| NetSimile | 0.345 | 0.017 | 5.740 | 0.400 | 3 |
| NetLSD | 0.462±0.047 | 0.924±0.001 | 77.830±72.704 | 0.389±0.014 | 5.900±5.878 |
| DeltaCon_0 | 0.254±0.010 | 0.701±0.002 | 25.960±0.260 | 0.334±0.004 | 12.000±0.000 |
| NPD | 0.222±0.039 | 0.407±0.051 | 22.459±8.234 | 0.572±0.074 | 9.800±1.874 |
| WL | 0.344 | 0.586 | 18.696 | 0.526 | 3 |
| LGCA-Sim | 0.902 | 0.954 | 226.364 | 0.258 | 6 |
| GCA-Sim | 0.929±0.026 | 0.959±0.027 | 764.841±555.807 | 0.233±0.071 | 4.500±1.354 |

Table 2. Clustering performance on 50 cities.

| Methods | Silhouette | Soft-Silhouette | CH | DB | Clusters |
|---|---|---|---|---|---|
| degreeJSD | 0.351 | 0.677 | 46.415 | 0.514 | 24 |
| NetSimile | 0.311 | 0.015 | 17.064 | 0.603 | 14 |
| NetLSD | 0.500±0.030 | 0.910±0.005 | 93.681±28.891 | 0.438±0.031 | 4.700±1.160 |
| DeltaCon_0 | 0.175±0.079 | 0.429±0.003 | 11.490±4.903 | 0.433±0.049 | 12.500±8.227 |
| NPD | 0.236±0.020 | 0.400±0.003 | 34.340±4.564 | 0.522±0.076 | 19.100±4.725 |
| WL | 0.303 | 0.562 | 5.558 | 0.393 | 3 |
| LGCA-Sim | 0.909 | 0.965 | 184.348 | 0.269 | 6 |
| GCA-Sim | 0.944±0.021 | 0.933±0.072 | 939.926±812.387 | 0.235±0.069 | 5.000±1.155 |

Table 3. Clustering performance on 50 districts.

The submodels under the GCA-Sim framework rank first on all metrics for both district and city samples. Their Silhouette scores exceed 0.9. By contrast, the best baseline, NetLSD, reaches only about 0.5. LGCA-Sim performs slightly worse than the ten other submodels in our framework, yet it still outpaces existing methods by a wide margin. Overall, models in the GCA-Sim framework generalize well. They distinguish similarity among urban spatial networks and produce tighter clusters with clearer separation.

We visualize the differentiable logic-gate network of the best submodel in Fig. 4(a).

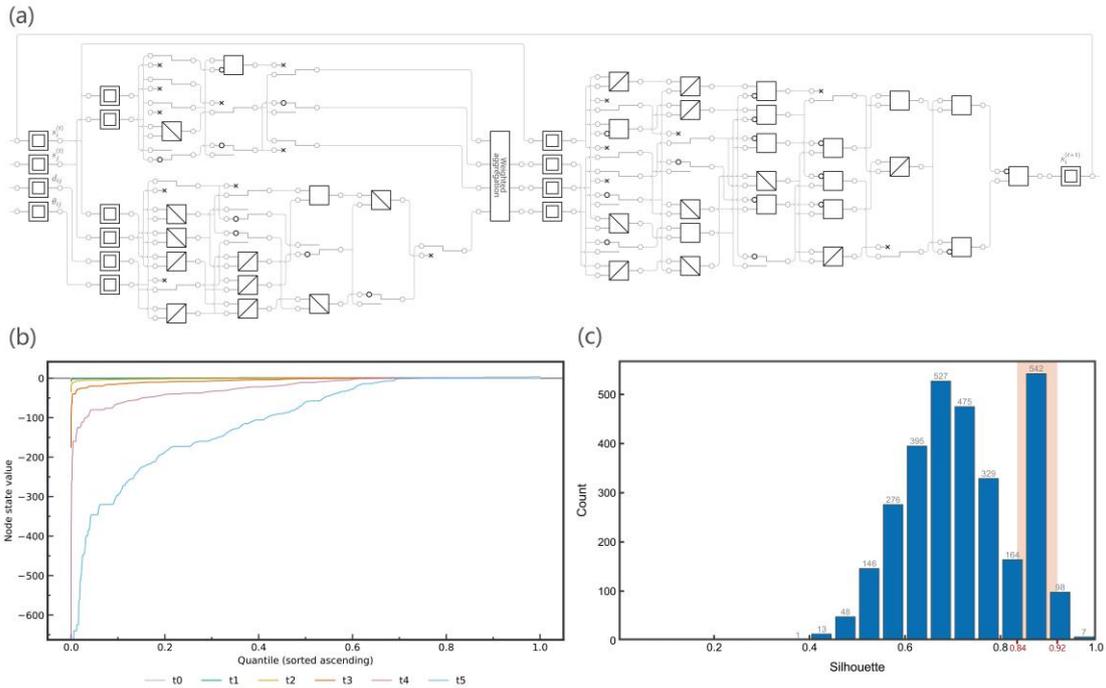

Fig. 4. Model visualization and performance analysis. (a) Differentiable logic-gate network of the best submodel. (b) Evolution of value distributions across iterations (quantile transformed horizontal axis). (c) Silhouette performance on the validation set during random initialization.

Two properties stand out. First, in the Fusion submodule routing dominates and transformation is secondary. Nonlinear gates are far fewer than in Attention and Update, which shows that Fusion tends to integrate information in a simple way. Second, in the Attention submodule the node state remains the main driver. Spatial features play a supporting role, with angle contributing more than distance. Using Shanghai as an example, we visualize the initial node states and the values across iterations with the best model in Fig. 4(b). The quantile contour curves grow steeper as iterations increase, which confirms the model's amplification effect on information.

We also summarize performance during random initialization across all parameter combinations on the validation set in Fig. 4(c). The median Silhouette is 0.71. A Gaussian mixture model (GMM) reveals two modes: a left mode with mean 0.68 that contains 80% of the data, and a right mode with mean 0.87 that contains 20%. The latter indicates that there are parameter settings that can amplify information interaction more strongly, which is the "network resonance" observed in this study.

## 4.2. Applications of the best submodel

### 4.2.1. Typology and Internal Consistency of Global Urban Networks

Fig. 5 reports similarity among the 50 cities in the analysis set and their internal consistency.

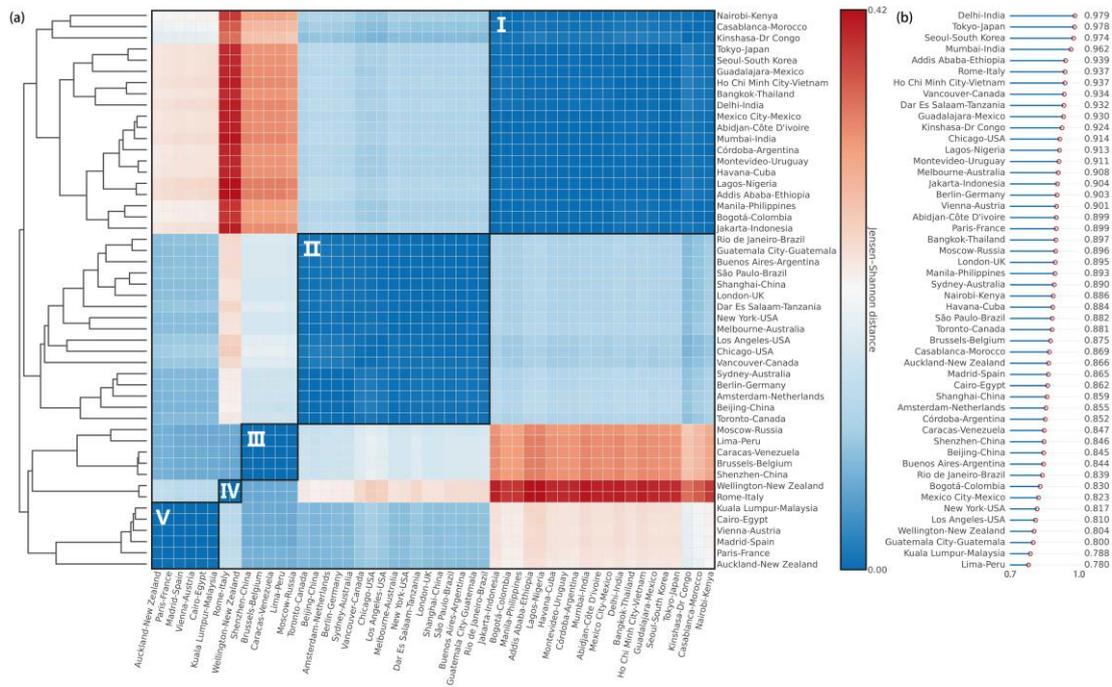

Fig. 5. Similarity and internal consistency of city-scale spatial networks. (a) Similarity matrix of 50 cities. (b) Internal consistency index for 50 cities.

The optimal partition has five clusters. In Cluster 1, which contains many Asian and African cities, internal consistency is high. These networks show bottom-up organic patterns and high network density. Cluster 2 includes cities in the Americas and Europe as well as Beijing and Shanghai. Internal consistency is lower. Networks are more regular and show clear planning. Cluster 3 has finger-like forms shaped by terrain and administrative boundaries. For example, Shenzhen and Lima are continuous along the coast, while mountains segment the opposite side into relatively independent parts. Cluster 4 contains only Rome and Wellington, whose networks are patchy: ties are strong within patches, but links between them are limited by terrain or historical protection and rely on only a few arterials. Cluster 5 combines a radial skeleton with a regular grid.

In general, cities with high internal consistency lack a strong global form, whereas cities with low internal consistency are mosaics of several clear organizational patterns.

We then evaluate similarity among 50 district-level units; Fig. 6 shows the results.

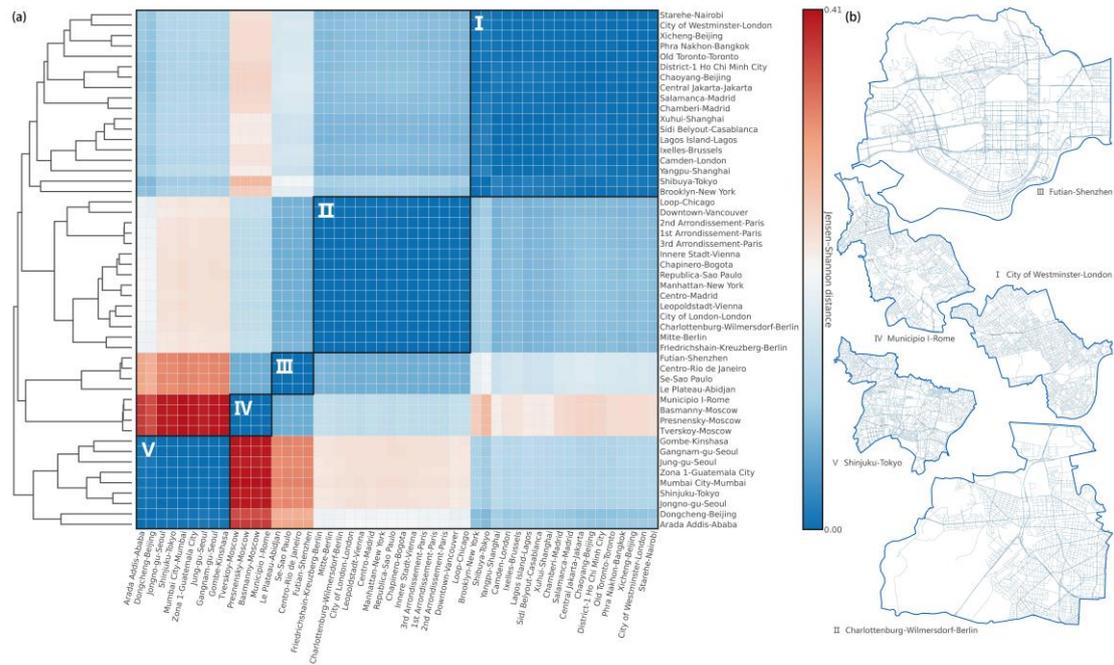

Fig. 6. Similarity of district-scale spatial networks. (a) Similarity matrix of 50 districts. (b) Representative medoid network of each cluster at a common scale.

The optimal partition again has five clusters. Clusters 1, 2, and 5 are mainly grid patterns with relatively homogeneous texture. Cluster 2 is the most regular: grid sizes are similar, roads are straight, and connectivity is strong in all directions, as in Manhattan. Cluster 1 is still orderly but shows variation in grid size and aspect ratio, with weaker connectivity on lower-rank roads. Cluster 5 is the most free-form. Grid variation is wide, and lower-rank connectivity is weakest. Clusters 3 and 4 reflect spatial barriers. Cluster 3 is coherent inside but bounded by expressways, waterways, or mountains that act as barriers. In Cluster 4, linear barriers such as major roads and rivers cross through the area, severing the low-rank networks and leading to the coexistence of multiple textures due to factors like heritage protection and development sequence.

### 4.2.2. Similarity and Continuity of Intra-urban Districts

Fig. 7 presents the results for Beijing.

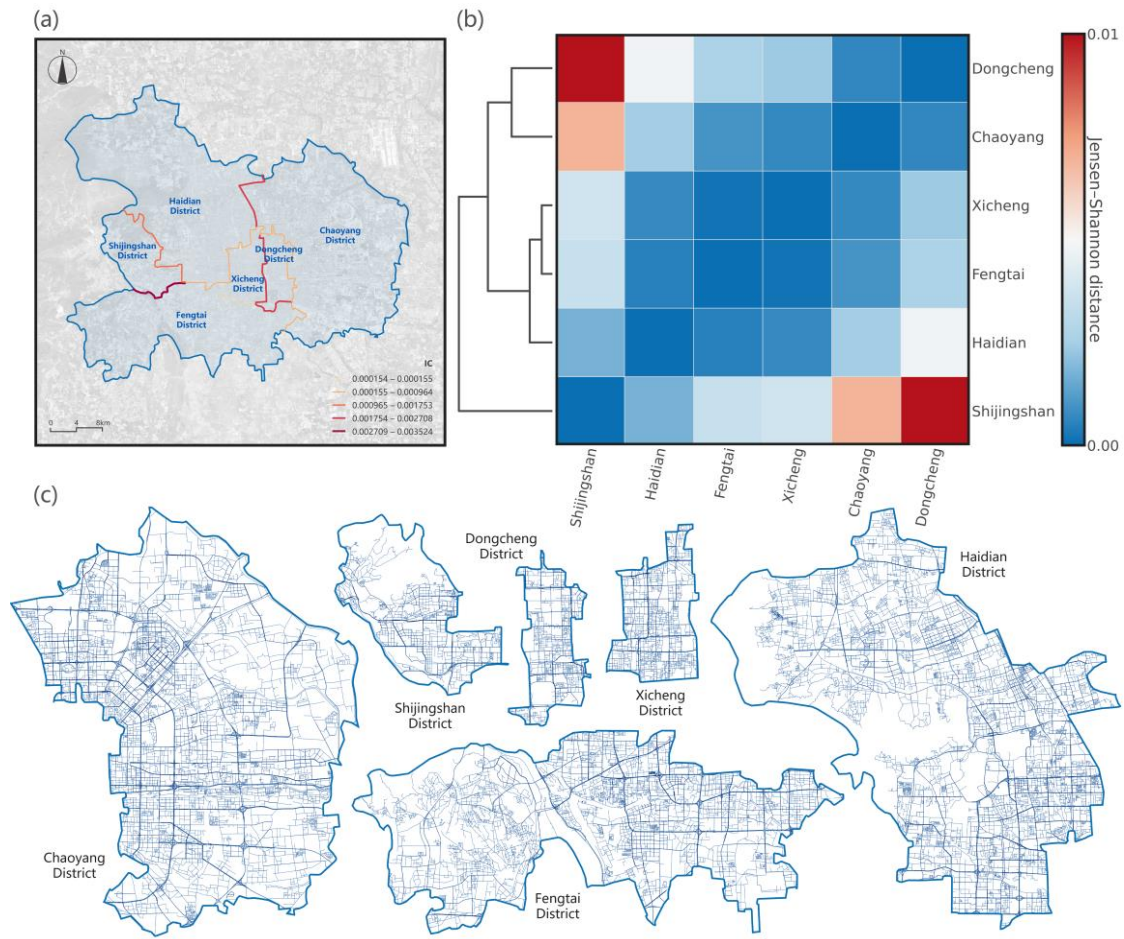

Fig. 7. Similarity among selected Beijing districts. (a) Continuity of boundary road networks between adjacent districts. (b) Similarity matrix for selected Beijing districts. (c) District road networks.

    Adjacent districts are generally more similar, while similarity falls with distance. For example, Shijingshan is closer to western districts and differs from the two eastern districts. Fig. 7(a) also shows that some neighboring districts differ markedly, which indicates breaks in texture. Although both sit at the urban core and retain traditional street fabric, Xicheng's connectivity is reduced by extensive waterways and large compounds, in contrast to Dongcheng's more continuous layout. Taking another example, the road network in Chaoyang is relatively regular. In contrast, Haidian is constrained by mountains in its west and includes large, later-developed areas in its northwest, resulting in greater heterogeneity. The largest gap is between Shijingshan and Fengtai, a difference that stems largely from the unique nature of Shijingshan. Encircled by various spatial barriers, it is relatively enclosed; furthermore, its core built-up area is a homogeneous grid. This combination of features makes it significantly different from the other districts.

    Fig. 8 reports the results for Shanghai.

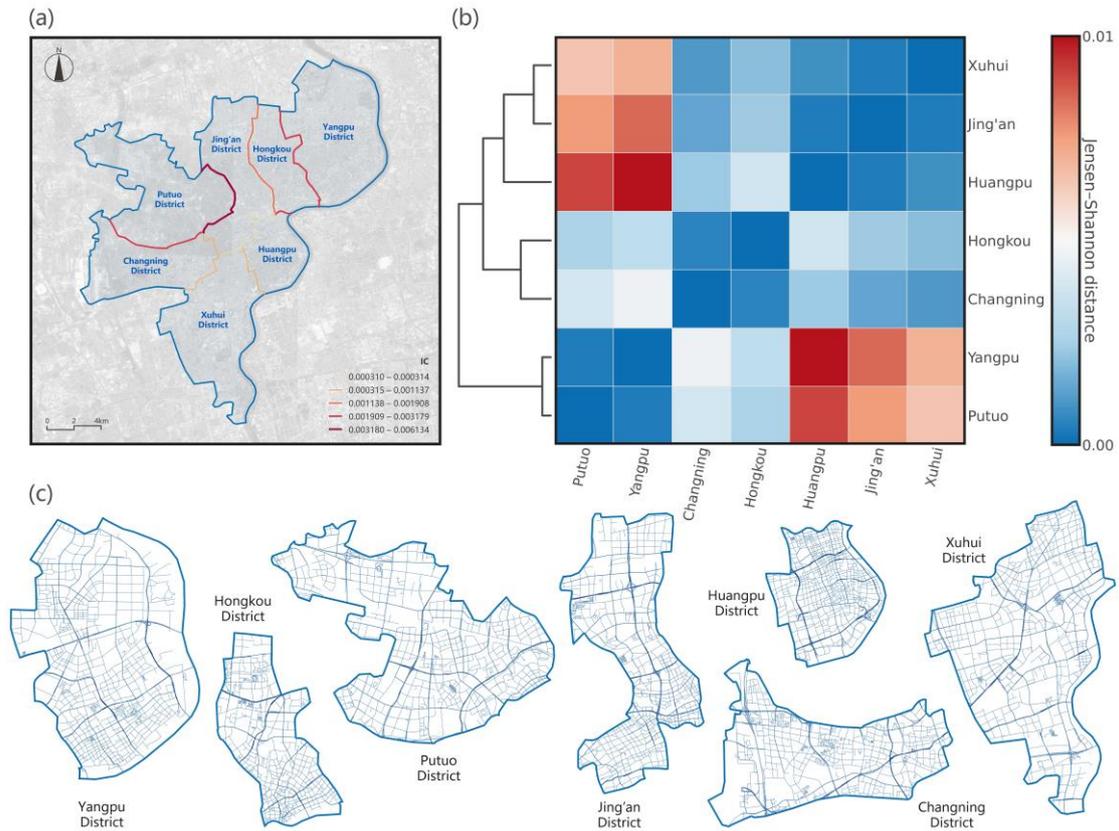

Fig. 8. Similarity among selected Shanghai districts. (a) Continuity of boundary road networks between adjacent districts. (b) Similarity matrix for selected Shanghai districts. (c) District road networks.

Three clusters are clear in the similarity matrix. The first is Huangpu–Xuhui–Jing'an. Huangpu, the historic core, has a denser network and retains much early urban fabric. Xuhui and Jing'an have undergone several rounds of administrative adjustment, which increases internal variation and produces slight differences from Huangpu. The second is Changning–Hongkou, a transition zone from the old city to the outer areas. These districts developed later than those in the first cluster and have medium density. The third is Putuo–Yangpu. Both lie on the outer edge of the central city and developed later. Planned grids are more evident, parcels are larger, and local streets are sparser, which makes them stand out from neighboring districts.

### 4.2.3. Correlations with external variables over iterations

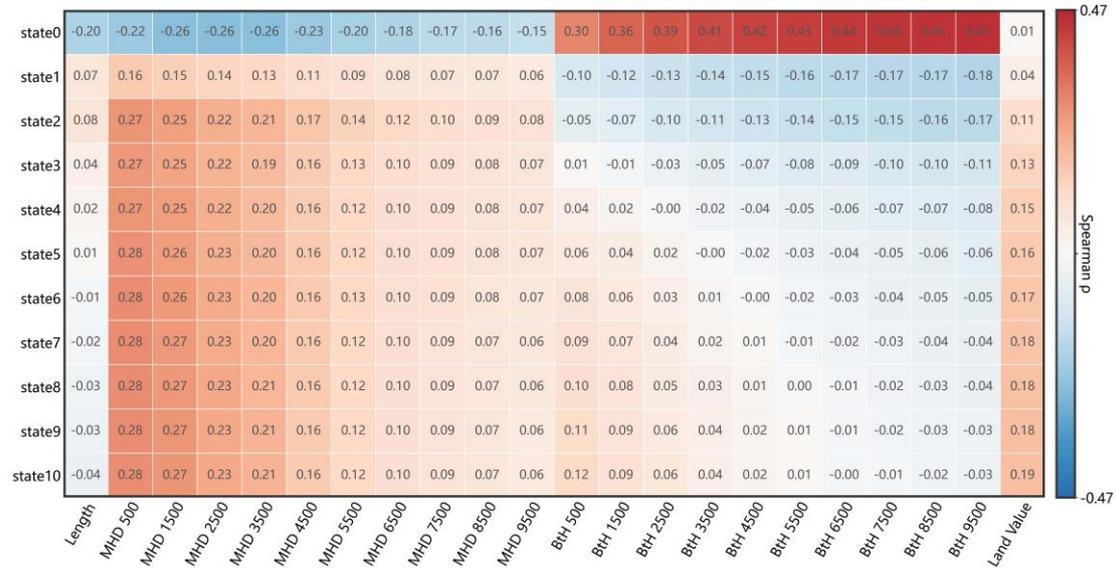

Fig. 9. Spearman correlations between degree-based iterative node states and external variables.

As shown in Fig. 9, under degree-driven dynamic propagation the agreement between node states and multiple variables increases steadily across iterations, though correlations remain weak. The largest change occurs in the first step, which marks a shift from self topology to neighborhood mixing. Correlation with mean road length changes little, which suggests a limited role for distance in this model. Correlation with MHD rises sharply in steps 1 to 3 and then plateaus; smaller analysis radii produce larger changes. Correlation with BtH increases only slightly overall, which reflects the method's focus on local diffusion rather than shortest-path betweenness. The most important finding is that the correlation with land value rises from near zero slowly and monotonically to a weak level, which suggests that iterations capture broader spatial structure and gradually build links to variables that reflect locational value.

## 5. Discussion

### 5.1. Performance of submodels under the framework

The submodels in our framework share several properties.

First, they fit the traits of urban spatial networks. Neighborhood updates reduce complexity to near linear, so the models can handle region-scale networks. The sampled variants of DeltaCon and NetLSD lower complexity but lose much information and depend on the sample, which hurts performance. In contrast, each GCA-Sim submodel yields deterministic similarity results. Dynamic information also simulates the evolution of initial states. It amplifies signals and helps capture subtle, latent differences, which suits road networks that are similar at first glance. The inclusion of distance and angle further matches the geographic nature of the networks.

Second, they have a high performance floor. Even the simplified LGCA-Sim leads the baselines by a large margin. With random initialization, most parameter settings still outperform existing methods. Diffusion by the graph Laplacian is a smoothing process, so the useful number of iterations is limited (Li et al., 2018). Some GCA-Sim submodels can also sharpen information, so iteration is not bounded in the same way and performance improves with a larger neighborhood. Whether smoothing or sharpening, propagation amplifies the influence of the signal on itself, which strengthens the ability to tell networks apart and raises the floor. In many cases the rule amplifies noise, but when it selectively amplifies structure that distinguishes network types, it triggers the "network resonance" observed here.

Third, they are label free. We do not predefine network types or assume that any pair is more similar than others. We rely only on internal clustering metrics. This reduces biases from experience and helps reveal prior knowledge hidden in similarity, for example the link between road-network form and development sequence in Shanghai. It also improves transferability across scales and new network types under the same evaluation standard.

Fourth, they are highly extensible. On the input side, the framework does not restrict the type or number of inputs, so one can match different scenarios. The models work even when the inputs are random, which reflects the interaction between structure and propagation. On the rule side, the update module can be replaced with preset or real-world propagation rules, such as pedestrian choice rules, and differentiable logic gates can support reverse discovery of the driving laws behind observed dynamics. One can also set the neighborhood range, run more iterations, record value distributions at intervals, or pair a center with nodes beyond k hops to compare broader neighborhoods.

## 5.2. Practical insights from the best submodel

The core value of exploring latent patterns is to uncover prior knowledge. The model places morphological features that we usually treat as different categories into a shared context, such as radial, patchy, or high-density grids. Clustering follows dynamic response rather than human Gestalt perception. Radial aggregation and uniform diffusion in dense grids both yield clear response signals. The clusters also cut across geography and income, as shown by the similarity between Tokyo and Guadalajara. This suggests that street-network form is a common language. We also find that planning-led networks show higher internal heterogeneity, while organically grown networks are more homogeneous. Planning is often a collage under many stages, actors, and objectives, which increases heterogeneity at the macro scale. Organic growth follows a stable and self-consistent growth logic over long periods and thus shows strong self similarity and internal consistency. The internal homogeneity detected by the model reflects consistency of generative rules rather than visual regularity alone.

A second finding is the emergence of socio-economic traits from topology. Initially, degree is merely an isolated node attribute. However, as each iteration aggregates information from a larger neighborhood, the node state evolves after several rounds into a comprehensive summary of the network's accessibility and complexity within its multi-hop neighborhood, thus simulating a process of locational value accumulation. This implies that street networks are not merely containers of activity; instead, form and function co-shape one another. Furthermore, it reveals a dynamic centrality that measures how much the structure promotes gains in flows of information, capital, and people, complementing traditional shortest-path centrality.

## 5.3. Limitations and Future Research

This study has three main limitations. First, we use only road networks and set degree as the initial state, so our results reflect topological similarity. We do not include economic or social data, which calls for tests in wider settings. Second, the rule set includes only basic nonlinear operators. Richer combinations are needed to approximate other nonlinear processes while keeping compression and interpretability. Third, our method learns bottom up. It remains hard to explain the root causes of spatial resonance in a closed form. For this reason we also provide the more interpretable LGCA-Sim.

Future work will proceed on three fronts. First, we will add socio-economic indicators, test whether the learned rules match real urban dynamics, and study how the similarity index relates to socio-economic variables, moving from morphological to functional similarity. Second, we will expand the operator set, add richer nonlinear forms and subgraph feature extraction, and push model compression and interpretability while keeping expressive power. Third, we will enhance the model's compatibility with three dimensional spatial networks, multilayer networks, and dynamic networks so that analyses of spatial continuity and internal consistency can enter micro-scale design and other settings, help designers quantify alignment with context, support organic continuity between old and new fabric, and improve the quality of the built environment.

## 6. Conclusions

This study began with an analogy to the Taylor expansion: if the distributions of change rates of all orders on a graph are similar, then the two spatial networks are similar. We therefore used the Laplacian operator, which reflects rates of change, as the evolution rule, yielding LGCA-Sim. It outperformed traditional models and showed the promise of dynamic information. We then asked whether a better rule exists. We made the parts of LGCA-Sim learnable and built the more flexible and extensible GCA-Sim. We found several submodels with stronger performance and confirmed the existence of propagation mechanisms that enlarge both similarity and dissimilarity between networks, that is, "network resonance." The best submodel identified latent patterns across scales in the analysis set. It also drove a steady rise in the correlation between iterative node states and land value.

Similarity is the foundation of generative design, as iterative optimization requires a clear direction, which in turn depends on a reliable judgment of whether the generated network matches the target. While existing methods struggle with this task, our framework offers a reliable basis for similarity evaluation in urban spatial networks, providing a foundation for generative design and in turn enabling the creation of virtual city scenarios. With large numbers of virtual scenes and controlled dynamics, we can clarify causal relations among urban factors (Pearl & Mackenzie, 2018) and support better decisions in urban governance. In sum, research on similarity in spatial networks underpins new urban science and digital twins and is key to understanding cities as complex systems.

## CRediT authorship contribution statement

**Peiru Wu:** Conceptualization; Methodology; Software; Validation; Formal analysis; Investigation; Data curation; Visualization; Writing – original draft; Writing – review & editing. **Maojun Zhai**: Data curation; Visualization; Writing – review & editing. **Lingzhu Zhang**: Supervision; Methodology; Writing – review & editing; Funding acquisition.

## Declaration of competing interest

The authors declare that they have no known competing financial interests or personal relationships that could have appeared to influence the work reported in this paper.

# Data availability

The data and code used in this study can be accessed at: https://github.com/PeiruWu0096/GCA-Sim